\documentclass[sigconf,nonacm]{acmart}

\usepackage{makecell}

\usepackage{amssymb}
\usepackage{pgfplots}
\pgfplotsset{compat=1.18}  
\usepackage{tikz}
\usepackage{array} 
\usepackage{array, tabularx, makecell}
\AtBeginDocument{%
  }

\setcopyright{cc}
\setcctype[4.0]{by}

\acmConference[Conference acronym 'XX]{Make sure to enter the correct
  conference title from your rights confirmation email}{June 03--05,
  2018}{Woodstock, NY}
\acmISBN{978-1-4503-XXXX-X/2018/06}

\makeatletter
\renewcommand\@secfont{\normalfont\bfseries\Large} 
\renewcommand\@subsecfont{\normalfont\bfseries\large} 
\renewcommand\@subsubsecfont{\normalfont\bfseries\normalsize} 
\makeatother

\begin{document}

\title{WebAssembly on Resource-Constrained IoT Devices: Performance, Efficiency, and Portability}

\author{Mislav Has}
\email{mislav.has@fer.hr}
\affiliation{%
  \institution{Faculty of Electrical Engineering and Computing}
  \city{Zagreb}
  \country{Croatia}
}
\author{Tao Xiong}
\email{tao.xiong@tietoevry.com
}
\affiliation{%
  \institution{Tietoevry Sweden Service AB}
  \city{Stockholm}
  \country{Sweden}
}
\author{Fehmi Ben Abdesslem}
\email{fehmi.ben.abdesslem@ri.se}
\affiliation{%
  \institution{Rise Research Institutes of Sweden AB}
  \city{Stockholm}
  \country{Sweden}
}
\author{Mario Kušek}
\email{mario.kusek@fer.hr}
\affiliation{%
  \institution{Faculty of Electrical Engineering and Computing}
  \city{Zagreb}
  \country{Croatia}
}

\renewcommand{\shortauthors}{}

\begin{abstract}
   The increasing heterogeneity of hardware and software in the Internet of Things (IoT) poses a major challenge for the portability, maintainability and deployment of software on devices with limited resources. WebAssembly (WASM), originally designed for the web, is increasingly recognized as a portable, secure and efficient runtime environment that can overcome these challenges. This paper explores the feasibility of using WASM in embedded IoT systems by evaluating its performance, memory footprint and energy consumption on three representative microcontrollers: the Raspberry Pi Pico, the ESP32 C6 and the nRF5340. Two lightweight WASM runtimes, WAMR and wasm3, are compared with the native C execution. The results show that while the native execution remains superior in terms of speed and energy efficiency, WASM offers acceptable trade-offs in return for cross-platform compatibility and sandbox execution. The results highlight that WASM is a viable option for embedded IoT applications when portability and security outweigh strict performance constraints, and that further runtime optimization could extend its practicality in this area.

\end{abstract}



\maketitle

\section{Introduction}

The Internet of Things (IoT) ecosystem is characterized by a large and growing diversity of hardware platforms, operating systems and development environments. While this heterogeneity enables a wide range of applications, it also introduces developers with significant challenges when it comes to maintaining and deploying software on resource-constrained devices. Traditional approaches often require platform-specific code or extensive rework to ensure compatibility and efficiency on each target. In this context, WebAssembly (WASM) has emerged as a promising technology for standardising development in the IoT landscape.

Originally designed for the web, WASM is a compact, binary instruction format that enables high-performance execution of code compiled from languages such as C, C++, and Rust. Its main strength lies in its portability: the same WASM binary can be executed on different platforms and architectures with minimal changes, given a compatible runtime environment is available. This opens up the possibility to develop once and deploy anywhere, which is a particularly compelling proposition for the fragmented and resource-constrained world of embedded IoT.

In recent years, the WebAssembly ecosystem has expanded beyond browsers into areas such as cloud computing, edge computing, and now embedded systems. This shift has been supported by the development of lightweight WASM runtimes that can run on microcontrollers and small IoT platforms. As a result, WASM is no longer limited to high-end environments and is being seriously considered as a unifying layer for portable and efficient embedded application development.

While WASM is increasingly promoted for its portability in IoT, there remains a lack of systematic evaluation of its performance and energy implications on resource-constrained microcontrollers. This paper addresses this gap by experimentally assessing the execution time, memory footprint, and energy consumption of representative WASM runtimes compared to native execution across multiple IoT devices. In doing so, it provides a quantitative analysis of WASM performance on low-power embedded platforms, examines the runtime-level differences between lightweight interpreters such as wasm3 and more feature-rich runtimes like WAMR, and discusses the trade-offs between portability, performance, and energy efficiency. The experimental design and methodology are based on prior work conducted in a thesis project in \cite{xiong2024performance}.

The paper is structured as follows: Section 2 provides an overview of existing research to establish the foundation and relevance of this study. Section 3 introduces the basic concepts of WebAssembly and explores its applicability to IoT environments. Section 4 describes the setup and methodology of the experimental environment. Section 5 presents and analyzes the experimental results. Finally, Section 6 concludes the paper by summarizing key findings and implications.
\section{Related Work}

There are several papers and projects investigating IoT devices and WASM.
Authors in \cite{10.1145/3593434.3593454} conducted a controlled experiment by compiling three benchmarking algorithms from four different programming languages (i.e. C, Rust, Go, and JavaScript) compiled to WASM and running these languages with two different WASM runtimes on a Raspberry Pi 3B. They identified C and Rust as a solid option for WASM projects working with low-performance hardware traditionally used in IoT devices.

The paper in \cite{9797106} has presented the possibility of combining WASM with embedded devices. They discuss the WASM runtimes that embedded devices can support and their structure.
Authors in \cite{eriksson2021containerizing} have conducted tests on a Raspberry Pi and show that there are many cases where a WASM runtime outperforms a similar Docker + C solution.
To address the challenge of running WebAssembly on resource-constrained IoT devices, the paper \cite{10.1145/3498361.3538922} introduces a lightweight runtime environment tailored for device-and cloud-integrated applications. Their approach supports Ahead-of-Time (AOT) compilation of WebAssembly on such constrained platforms, employing memory reduction techniques and compile-time safety checks to ensure secure sandboxed execution. The results show that this approach reduces memory footprint by up to 84.8x and improves energy efficiency by 1.2x to 4.9x compared to existing AOT runtimes.
Moreover, authors in \cite{wagner2023energy} conducted a controlled experiment in which they tested WASM binaries generated from C, Rust, Go, and JavaScript on a Raspberry Pi 3B using two different runtimes. Their results show that the choice of source language has a significant impact on both energy consumption and execution performance, while the runtime environment plays a less decisive role. In particular, C and Rust performed better than other languages, while JavaScript compiled with Javy showed poor efficiency. 

Further studies, such as \cite{electronics13203979}, have explored hardware-assisted WebAssembly execution for embedded systems, demonstrating that dedicated acceleration units can substantially improve throughput and efficiency on low-power processors.
In addition, \cite{10205816} provided an extensive review of WebAssembly beyond the web, examining its performance characteristics, runtime constraints, and optimization opportunities in edge and resource-constrained environments.

While these studies have advanced the understanding of WebAssembly performance on single-board computers and moderately constrained platforms, few have focused specifically on ultra-constrained microcontrollers typical of IoT end devices.
This paper differs from prior work in three main ways:

(1) It extends the experimental scope to multiple low-power microcontrollers (Raspberry Pi Pico, ESP32 C6, and nRF5340) to capture performance variations across architectures;

(2) It directly compares lightweight and feature-rich WASM runtimes (wasm3 and WAMR) under identical workloads to identify trade-offs in execution time, energy consumption, and memory usage;
    
(3) It documents practical challenges encountered when deploying and executing WASM on these devices, providing implementation insights useful for future research and development.

By addressing these aspects, this study complements existing research and contributes new empirical data and methodological guidance for evaluating WASM in resource-constrained IoT environments.

\section{WebAssembly Background and IoT Integration }

WebAssembly is a binary instruction format for a stack-based virtual machine.
It was developed as a portable target for compiling high-level languages such as C, C++, and Rust, so that these languages can be used on the web for both client and server applications. While JavaScript remains the only programming language natively supported by web browsers, WASM fills this gap by providing a solution for running safe, fast, and portable low-level code on the web. Therefore, it works alongside JavaScript and often serves as a complementary technology. to improve performance and enable the execution of more complex tasks. The basic design goals of WASM are as follows \cite{fi15080275}:

\begin{itemize}
    \item Fast transfers over the Internet: Considering JavaScript is a plain text format, it cannot achieve the compact file sizes that WASM can.
WASM is designed for compact and load-time saving binary representation, which reduces load time and saves bandwidth.
    \item Security: Security is crucial for web technologies because the code comes from untrusted sources. WASM has a memory-safe execution environment (sandbox) that can be isolated from the host runtime. Other web technologies such as Javascript use a managed language runtime, which can have a negative impact on performance due to the overhead of ensuring security. Essentially, WASM achieves a balance between security and performance by minimizing the performance degradation normally associated with secure execution.
    \item  Portability: The structure of the WASM binary format is designed to run efficiently on different operating systems and instruction set architectures. When a program is compiled into WASM bytecode, it can be distributed and executed in different environments as long as there is a WASM runtime environment is available.
    \item Open and debuggable: There are several tools and API support for WASM. WASM can also be printed out in a textual format for debugging, testing, experimenting, optimizing, learning, teaching and writing programs by hand. The format allows developers to view the source code of WASM modules on the web, making it more accessible for various development tasks.
    \item Part of the open web platform: Another design goal of WASM is to maintain the versionless and feature-tested nature of the web. Its modules can be called into and out of the JavaScript context and access browser functionality.
\end{itemize}

While the Internet of Things is driving technological progress in many areas, it still faces some major challenges. One major issue is scalability and diversity. IoT systems consist of numerous devices and infrastructures, each using different architectures, protocols, and data formats. This leads to major difficulties in ensuring interoperability and seamless integration in large-scale, heterogeneous environments. Security and data protection are also pressing concerns. IoT networks are often the target of attacks, and although there are various security measures in place, these are often accompanied by performance degradation. Energy efficiency is another critical challenge. Many IoT devices are battery-powered, and factors such as unnecessary data transfers or constant wireless usage can quickly consume energy, making power optimization essential.

Finally, limited hardware resources are also a constraint. Tiny IoT devices often lack the processing power and memory to run full operating systems. While lightweight real-time operating systems can be used, they do not offer the same level of flexibility or isolation.

WASM offers some features and benefits that are notably impactful for some
of the mentioned IoT challenges:

\begin{itemize}
    \item Flexibility: WASM was developed as a platform-independent bytecode that increases flexibility in IoT development. It allows the same code to run on different devices and platforms, eliminating the need for platform-specific customization. This feature can improve the efficiency of new technology development, which is critical for rapidly evolving IoT ecosystems \cite{fi15080275}.
    \item Security and data protection: The usual, widely used memory protection methods are not available for low-end microcontrollers. However, by implementing WASM on these devices, the problem can be solved without the need for virtual memory and extensive hardware resources. The reason for this is that WASM provides an abstraction layer that allows applications to run in a sandbox environment so that the code can be executed in a controlled environment. This can help protect the host system from potentially malicious operations \cite{9797106}.
    \item Compactness and Energy Efficiency: WASM has a binary instruction set, which can be directly processed by the device, saving energy for resource-intensive operations. This efficiency feature can lead to lower energy consumption for CPU usage, especially for battery-powered IoT devices.
    \item  Ahead of Time compilation: Some WASM runtimes support Ahead of Time compilation, which takes the WASM bytecode and produces machine code for the target CPU/MCU type. This is very useful in the context of tiny IoT devices, which may have limited CPU and memory to perform Just in Time compilation, as we typically do in the cloud or on the desktop \cite{10.1145/3062341.3062363}.
\end{itemize}

There are a variety of WASM runtimes for web and non-web \cite{10.1145/3714465}, which have features \cite{appcypher2024awesome}, such as performance optimization, portability, and so on. In this study, we focus on runtimes that support IoT devices.  Table \ref{tab:runtimes} shows the comparison of the five most popular WASM runtimes. Each of these runtimes has different features and capabilities tailored to different development needs.

\begin{table}[ht!]
\centering
\caption{The most popular WASM runtimes.}
\label{tab:runtimes}  
\begin{tabularx}{\columnwidth}{|>{\centering\arraybackslash}X|c|c|c|c|}
\hline
\textbf{\makecell{Wasm \\ Runtime}} &
\textbf{\makecell{Standalone \\ Interpreter}} &
\textbf{JIT} &
\textbf{AOT} &
\textbf{GitHub Stars} \\
\hline
\makecell{wasmer\cite{wasmer2024}} & \checkmark & X & \checkmark & 19.9k \\ \hline
\makecell{wasmtime\cite{bytecode2024wasmtime}} & \checkmark & X & \checkmark & 16.6k \\ \hline
WasmEdge\cite{wasmedge2024runtime} & \checkmark & X & X & 9.6k \\ \hline
wasm3\cite{wasm32024} & \checkmark & \checkmark & X & 7.6k \\ \hline
WAMR\cite{bytecode2024wamr} & \checkmark & \checkmark & \checkmark & 5.4k \\ \hline
\end{tabularx}
\end{table}

Due to its robust support for Just-in-Time (JIT) and Ahead-of-Time (AOT) compilation, WAMR is the first choice to improve performance and flexibility. JIT compilation converts the WASM code into machine code at runtime to improve performance. AOT compilation compiles WASM code into native machine code before execution, reducing startup times. In addition, these runtimes have a large community, resulting in better support, frequent updates, and a more extensive ecosystem, which is crucial for academic and practical implementations.
WAMR has more comprehensive support for all the features, which makes it versatile for comprehensive applications. Additionally, it supports various IoT platforms, which makes them ideal for research on lightweight, cross-platform runtime environments in resource-constrained devices.
Compared with other runtimes, wasmedge and wasm3 offer a better balance between integration capabilities and runtime efficiency. Specifically, wasm3 focuses more on speed with selective compilation features, while wasmedge pays more attention to interoperability.

\section{Implementation}

To evaluate the execution of WASM on resource-constrained devices, we conducted experiments with two runtimes: WAMR and wasm3. All measurements were performed with interpretation enabled, not ahead-of-time (AOT) or just-in-time (JIT) compilation, to better reflect the capabilities and constraints typical of embedded environments. Selection of the runtime for each platform was based on its availability, maturity, and compatibility. WAMR was chosen for the Raspberry Pi Pico, ESP32 C6, and nRF5340 due to its flexibility and Zephyr integration, while wasm3 was used for the Pico and ESP32 C6 due to its lightweight and fast interpretation capabilities. To ensure a meaningful comparison between different environments and languages, we tested various computational algorithms implemented in both C and Rust. These included a bubble sort function to represent control-flow intensive workloads and a CRC-16 checksum calculation to simulate bitwise and performance-critical operations. The choice of these workloads reflects typical patterns in embedded applications, illustrating the trade-offs developers face between control-flow complexity and computational intensity. These examples allowed us to assess runtime behaviour, performance, and compatibility under different code structures and computational workloads. The following subsections describe the setup and execution of the experiments for each platform. 

\subsection{Experiments Execution on Pico}

\subsubsection{Wasm3 experiments}

\begin{figure}[h]
    \centering
    \includegraphics[width=1\linewidth]{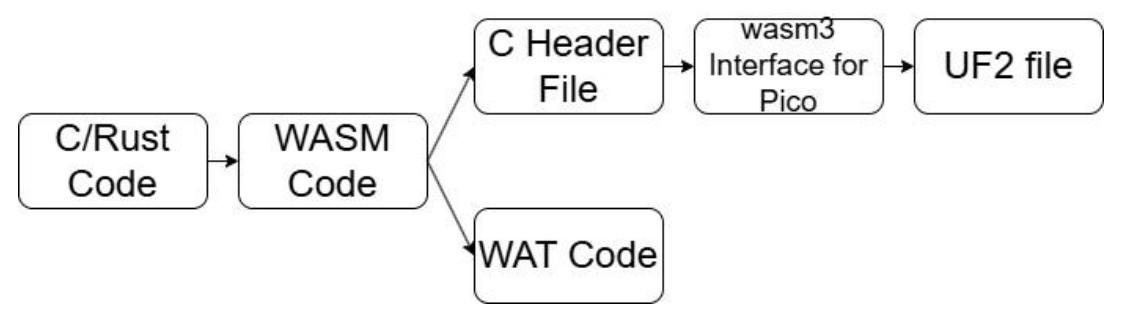}
    \captionsetup{justification=centering}
    \caption{Wasm3 execution on Pico.}
    \label{fig:wasm3-pico}
\end{figure}

The entire process of running wasm3 on Pico is shown in Figure \ref{fig:wasm3-pico}. At the beginning, there is a C or Rust code in the project, which is used as source code for the generation of
WASM code. Next, a bash script is executed. This script compiles the C or
Rust code into WASM code with the help of the Clang compiler from the WASI SDK.
At the same time, it generates a WAT file (WebAssembly Text Format) to enable better analysis of the WASM code. Subsequently, xxd is used to convert a binary file (WASM file) into a C header file format. The xxd is then used to convert the binary WASM file into a C header format. A separate
C file as an interface to connect the WASM code to the C
code embedded in Pico. The generated C header file is integrated into this interface code, and the final output is a .uf2 file that can be flashed directly to the Pico and can be transferred.

\subsubsection{WAMR experiments}

\begin{figure}[h]
    \centering
    \includegraphics[width=1\linewidth]{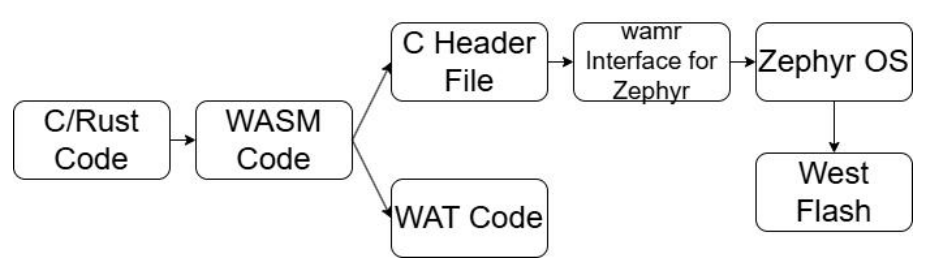}
    \captionsetup{justification=centering}
    \caption{WAMR execution on Pico.}
    \label{fig:wamr-picoo}
\end{figure}

The entire process of running WAMR on Pico is shown in Figure \ref{fig:wamr-picoo}.
First, C or Rust code is compiled into WASM code. Then, a bash script is executed to generate WAT code and a C header file. Finally, West, a tool from Zephyr, is used to build and flash the project in Pico. For the C and Rust projects, the same examples are selected that are used for Pico with wasm3.

\subsection{Experiments Execution on ESP32 C6}

\subsubsection{Wasm3 experiments:}

Running wasm3 with ESP32 C6 follows a process similar to running wasm3 with Pico.
However, there is direct support from the wasm3 community for ESP32 C6,
which allows running wasm3 on the ESP32 C6 without additional platform
support. Consequently, we can utilize the wasm3 ESP32 C6 interface
directly, rather than developing a custom interface as was necessary for the
Pico.

\subsubsection{WAMR experiments}

As mentioned previously, after adding support for ESP32 C6 to WAMR, it
is now possible to run WASM code directly on ESP32 C6. Consequently,
the process is similar to WAMR and Pico, but without using Zephyr. First, C or
Rust code is compiled into the corresponding WASM code. Next, it will generate
WAT file for analysis and a C header file. Moreover, the header file will then
be integrated into the WAMR interface file for ESP32 C6. Finally, ESP tools,
primarily Python-based, are used to build the project, flash the code, and
monitor the output.

\subsection{Experiments Execution on nRF5340}

\subsubsection{WAMR experiments}

Because WAMR does not have direct support for the nRF5340 board, but
Zephyr does, the process follows the same steps as the WAMR setup on the Pico, as shown in Figure \ref{fig:wamr-picoo}.

\section{Performance and energy consumption analysis}

To assess the runtime efficiency of WASM on constrained embedded systems, we performed a detailed evaluation of the performance and energy consumption of three microcontrollers: the Raspberry Pi Pico, ESP32 C6, and nRF5340. In these experiments, WAMR and wasm3 runtimes were directly compared to native C execution, which serves as a baseline across all devices.

We tested the performance and energy behavior using code written in both C and Rust. To challenge the runtimes, we implemented several computational workloads: a control-intensive algorithm using bubble sort with a varying number of integers, 100 and 1000, and a CRC-16 algorithm with 100 integers as a data set. Although these benchmarks are simple, they reflect common classes of embedded workloads and provide reproducible test cases that can be easily replicated by others. The results shown in the following figures reflect the consistent pattern observed across all tests: native implementations were the most efficient in terms of energy and execution time, while WASM runtimes, especially wasm3, offered competitive performance at a higher but manageable resource cost.

\subsection{Energy consumption measurement}

\begin{figure}[h]
    \centering
    \includegraphics[width=1\linewidth]{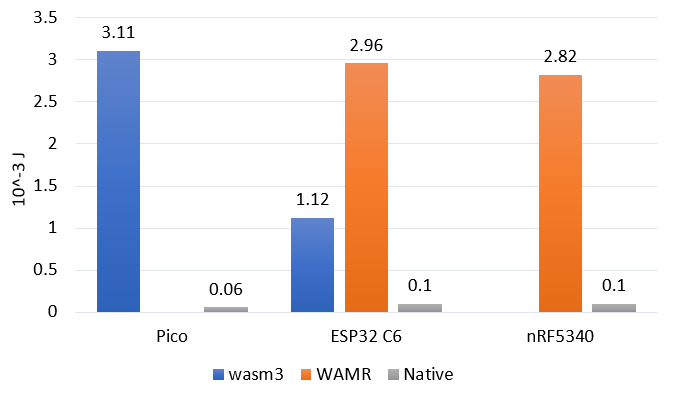}
    \captionsetup{justification=centering}
    \caption{Energy consumption of devices with Bubble Sort algorithm.}
    \label{fig:consum}
\end{figure}

The energy consumption of a device can be determined by multiplying the average power by the duration of the interval during which this average power is calculated. This relationship is expressed by the formula: $E = P_{\text{average}} \times T_{\text{execution}}$.
In all experiments, we put the device to sleep for 5 seconds before and after the execution of the benchmark algorithms. In this case, there is a short increase in energy consumption caused by the benchmark algorithm. We repeat the same steps several times to obtain reliable results. With this approach, we can calculate the energy consumption by determining the average power consumption during the active phase and multiplying it by the execution time of the function. The product of these values gives the total energy consumption for the execution of the function. All measurements were repeated several times to minimize the effect of variability and to confirm consistency across runs. 
The formula is used for all three tiny IoT devices in the project to easily compare the results. Therefore, we measure the average current and voltage supplied, as the power can be calculated using the formula P = V × I.
Then we multiply these values by the execution time to determine the energy consumption.
In this context, the Power Profiler Kit II \cite{nordic2024ppk2}, also known as PPK2, is used to measure the power consumption of all three devices. PPK2 is a stand-alone device capable of measuring and optionally delivering currents in the range of sub-microamperes up to 1 ampere. It can be used with all Nordic
Development Kits (DKs) as well as with external hardware. With the help of the PPK2, we can monitor the current in real time and determine the peak current value. The PPK2 operates at a high sampling rate, which ensures that short-term peaks in current consumption are accurately detected.
The following graph in Figure \ref{fig:consum} shows the energy consumption of the Bubble Sort algorithm (100 integers, implemented in C) on three microcontrollers.

 The native C implementation consistently has the lowest energy consumption across all platforms. The values range from 0.06 mJ on the Pico to 0.1 mJ on the ESP32 C6 and the nRF5340. In contrast, WebAssembly runtimes have a significantly higher energy consumption. On the Raspberry Pi Pico, wasm3 consumed 3.11 mJ. On the ESP32 C6, wasm3 consumed 1.12 mJ, while WAMR consumed 2.96 mJ. On the nRF5340, only WAMR was supported and the energy consumption was measured at 2.82 mJ. The WAMR runtime shows a relatively constant energy consumption on both the ESP32 C6 and the nRF5340 (2.96 mJ and 2.82 mJ respectively), indicating stable behavior across different architectures. However, wasm3 exhibits a significant difference between the platforms. Energy consumption drops significantly from the Pico (3.11 mJ) to the ESP32 C6 (1.12 mJ). This difference can be attributed to the underlying microarchitecture or the hardware power management features. This emphasises an important trade-off: WASM runtimes offer cross-platform compatibility, but at the cost of higher energy consumption on devices with limited resources.

\subsection{Memory Footprint Measurement}

The methods of memory footprint measurement are different for the various devices, as they use different methods for flashing the firmware.
\subsubsection{Pico:}
The static memory footprint for the Pico consists of the size of the UF2 file and the RAM usage during program execution. The size of the UF2 file indicates the memory allocated in the flash memory of the Pico, while the RAM usage refers to the memory required by the program while running.
First, the source code is compiled to create a UF2 binary file for flashing the firmware to the Pico. The size of the UF2 file, which represents the static memory in the microcontroller’s flash memory, is recorded. During program
 execution, RAM usage is monitored using profiling tools or development environment utilities. The maximum RAM usage is recorded in order to determine the maximum dynamic memory requirement. 

\subsubsection{ESP32 C6:}

The IDF toolchains include the ‘idf.py size‘ command, which provides a comprehensive summary of memory usage that includes the dimensions of text and data sections. Therefore, this command can be used directly to quantify the memory requirements of the project. 
\subsubsection{nRF5340:}
Upon completion of the project build, Zephyr displays a detailed summary of the project’s memory usage. For the purposes of this study, which requires an assessment of static memory usage, the cumulative total of FLASH and RAM regions utilized is considered. This comprehensive measurement includes all allocated memory segments and provides a clear and precise evaluation of the memory resources consumed by the project.

\begin{figure}[h]
    \centering
    \includegraphics[width=1\linewidth]{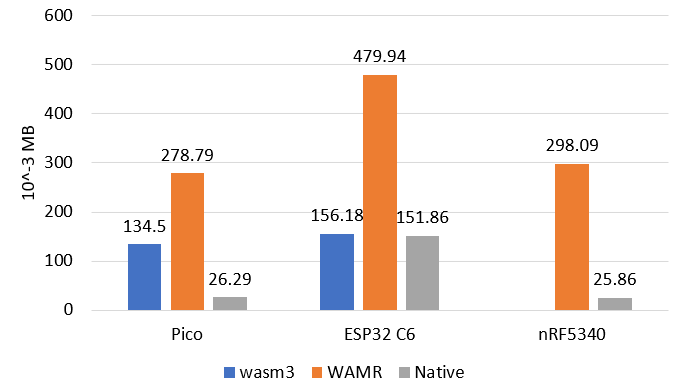}
    \captionsetup{justification=centering}
    \caption{Memory footprint of devices with Bubble Sort algorithm.}
    \label{fig:wamr-pico}
\end{figure}

The memory footprint comparison across platforms for the 100-integer bubble sort task shows consistent trends in runtime memory requirements. The native implementations occupy the smallest amount of memory on all three devices, confirming their minimal overhead. Among the WebAssembly runtimes, WAMR consistently shows the highest memory consumption, with values of about 279 KB on the Pico, almost 480 KB on the ESP32 C6, and 298 KB on the nRF5340. This high consumption reflects the additional memory required to manage the WASM environment and the runtime abstraction layers.
In contrast, wasm3 has a more compact memory profile, especially on the Pico (about 134 KB) and the ESP32 C6 (156 KB), and is not available on the nRF5340 in this test. The relatively smaller gap between wasm3 and native code suggests that it may be a more memory-efficient choice for constrained systems, although it is still significantly heavier than native execution.

\subsection{Execution Time Measurement }

\begin{figure}[h]
    \centering
    \includegraphics[width=1\linewidth]{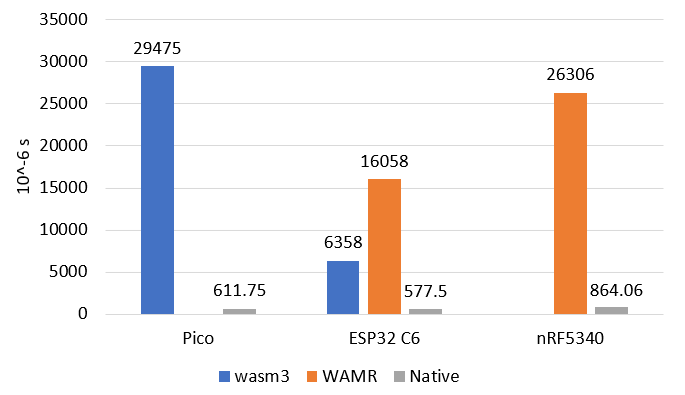}
    \captionsetup{justification=centering}
    \caption{Execution time of devices with Bubble Sort algorithm.}
    \label{fig:exec}
\end{figure}

The execution time measured refers to the time required by the benchmark algorithms for their execution. For WASM code, it is defined as the interval between the start of the execution of the function by the WASM runtime environment and the termination of the function. For native code, the execution time is measured from the time the function is executed until it is completed. In addition, different IoT devices and WASM runtimes have different libraries. Therefore, each device uses its own method to measure the execution time. The comparison of execution times for the 100-integer bubble sort algorithm on the Pico, ESP32 C6, and nRF5340 platforms is visible in Figure \ref{fig:exec}.

The native code achieves the fastest execution times on all devices: 611.75 µs (Pico), 577.5 µs (ESP32 C6), and 864.06 µs (nRF5340), demonstrating the advantage of running directly on the hardware with minimal overhead.
WASM runtimes, while naturally introducing some overhead due to their abstraction layer, demonstrate encouraging performance given their runtime features and flexibility. Wasm3 consistently delivers solid execution times, for example, 6,358 µs on the ESP32 C6 compared to 577.5 µs natively, making it a practical choice when balancing performance and runtime functionality. While WAMR has higher execution times (e.g., 29,475 µs on Pico), it offers a rich feature set and supports a wider range of development scenarios, making it a good choice for more complex or modular applications.

\subsection{Discussion}

Although native execution consistently outperforms WASM runtimes in raw speed and energy efficiency, the results highlight several scenarios in which WASM offers distinct advantages for IoT systems. These advantages lie mainly in portability, safety, and maintainability across heterogeneous devices rather than pure performance.

In heterogeneous IoT deployments, devices often differ in architecture and operating systems, complicating firmware updates and maintenance. WASM mitigates this by providing a portable, architecture-independent execution layer that allows the same bytecode to run on multiple platforms with minimal modification. This feature simplifies over-the-air (OTA) updates and enables uniform code deployment across diverse hardware. Moreover, WASM’s sandboxed execution model isolates untrusted code, improving security for devices that must execute third-party or dynamically loaded modules, such as industrial gateways or shared sensor nodes. Its modularity also allows IoT gateways to dynamically load or update functionalities without reflashing firmware, supporting adaptive and long-lived systems.

The comparative analysis of runtimes shows that wasm3 and WAMR serve complementary purposes. wasm3 is a lightweight interpreter optimized for constrained environments, offering fast startup, low memory usage, and predictable behavior. These qualities make it ideal for simple, battery-powered end devices or control tasks where deterministic execution and minimal footprint are critical \cite{10.1145/3714465}. In contrast, WAMR provides a richer feature set, AOT and JIT compilation, multi-threading, and integration with operating systems like Zephyr \cite{wasm_micro_runtime_docs}. These capabilities make WAMR better suited for edge or gateway devices that can leverage runtime extensibility and partial compilation to improve performance despite higher overhead. However, this study focused exclusively on the interpretation mode for both runtimes, which reflects the operational reality of highly resource-constrained IoT hardware. The omission of JIT and AOT execution modes represents a limitation, as these approaches can substantially improve performance and reduce energy consumption in more capable devices.

Finally, wasm3 prioritizes minimalism and efficiency for resource-limited devices, while WAMR emphasizes flexibility and integration for more capable embedded systems. Although both runtimes trail native execution in raw performance, their security, portability, and maintainability make them strong candidates for scalable, cross-platform IoT development.

\section{Conclusion}

The experimental results show a consistent trend: native implementations outperform WASM on key performance metrics, including execution time, energy consumption and often memory usage.  However, performance varied significantly between WASM runtimes. Lighter-weight runtimes, such as wasm3, generally executed algorithms faster and with a lower memory footprint than heavier runtimes such as WAMR. These differences illustrate how runtime design and optimization significantly affect efficiency, especially on resource-constrained hardware.
While certain platforms struggled to run WASM due to limited memory capacities or slower processing capabilities, others proved to be more powerful and better supported. Devices with larger memory capacity and broader runtime compatibility were generally better suited for WASM deployment, especially for more complex or memory-intensive applications.

Despite its current limitations in performance and energy efficiency, WASM remains a compelling option in scenarios where portability, security and cross-platform development are priorities. For applications where resource conservation or low-latency execution is critical, native code remains the preferred approach. However, as WASM runtimes continue to mature and embedded hardware evolves, the trade-offs become more favorable. This study concludes that while WASM is not yet a complete replacement for native execution in constrained environments, it is an increasingly practical solution, especially when development flexibility and platform independence are desired. These findings can also guide developers in selecting the most suitable WASM runtime for their specific hardware and application requirements, balancing efficiency, memory usage, and feature support.

Future work will explore a broader range of WASM runtimes to better assess their suitability for diverse IoT use cases, especially as the technology continues to mature. This includes evaluating additional classes of resource-constrained devices to gain deeper insight into the practical challenges of deploying and executing WASM in such environments. Future studies should also conduct a systematic comparison of interpreted, AOT, and JIT execution modes to quantify their respective trade-offs in execution speed, memory footprint, and power efficiency under varying IoT workloads. Another promising direction is the integration of real-time operating systems (RTOS) with WASM, which could enable more responsive and capable embedded applications. As both WASM runtimes and embedded hardware continue to evolve, these efforts will help clarify the role of WASM in the next generation of IoT systems.

\section{Acknowledgments}
This work has been supported by the Horizon Europe WIDERA program under the grant agreement No. 101079214  (AIoTwin).

\bibliographystyle{unsrt}
\bibliography{references.bib}

\appendix

\end{document}